\documentclass[runningheads]{llncs}
\usepackage[T1]{fontenc}
\usepackage{graphicx}
\usepackage{amsmath}
\usepackage{amssymb}

\begin{document}
\title{Topological and Graph Theoretical
Analysis of Dynamic Functional
Connectivity for Autism Spectrum Disorder}
\titlerunning{Topological and Graph Analysis of Autism fMRI}
\author{Yuzhe Chen\inst{1}
% \orcidID{0000-1111-2222-3333}
\and
Dayu Qin\inst{2}
% \orcidID{1111-2222-3333-4444}
\and
Ercan Engin Kuruoglu\inst{3}\thanks{Corresponding author}
% \orcidID{2222--3333-4444-5555}
}
\authorrunning{Y. Chen et al.}
\institute{Tsinghua-Berkeley Shenzhen Institute, Tsinghua University, Shenzhen 518055, Guangdong, China \email{chen-yz22@mails.tsinghua.edu.cn} \and
Tsinghua-Berkeley Shenzhen Institute, Tsinghua University, Shenzhen 518055, Guangdong, China
\email{tdy22@mails.tsinghua.edu.cn} \and
Tsinghua-Berkeley Shenzhen Institute, Tsinghua University, Shenzhen 518055, Guangdong, China
\email{kuruoglu@sz.tsinghua.edu.cn}}
\maketitle              % typeset the header of the contribution
\begin{abstract}
% The abstract should briefly summarize the contents of the paper in
Autism Spectrum Disorder (ASD) is a prevalent neurological disorder. However, the multi-faceted symptoms and large individual differences among ASD patients are hindering the diagnosis process, which largely relies on subject descriptions and lacks quantitative biomarkers. To remediate such problems, this paper explores the use of graph theory and topological data analysis (TDA) to study brain activity in ASD patients and normal controls. We employ the Mapper algorithm in TDA and the distance correlation graphical model (DCGM) in graph theory to create brain state networks, then innovatively adopt complex network metrics in Graph signal processing (GSP) and physical quantities to analyze brain activities over time. Our findings reveal statistical differences in network characteristics between ASD and control groups. 
Compared to normal subjects, brain state networks of ASD patients tend to have decreased modularity, higher von Neumann entropy, increased Betti-0 numbers, and decreased Betti-1 numbers. These findings attest to the biological traits of ASD, suggesting less organized and more variable brain dynamics. These findings offer potential biomarkers for ASD diagnosis and deepen our understanding of its neural correlations.
\keywords{Autism \and Dynamic functional connectivity \and Graph signal processing \and Topological data analysis \and fMRI
% \and Ollivier-Ricci curvature \and Laplacian eigenvalue.
% First keyword  \and Second keyword \and Another keyword.
}
\end{abstract}
\section{Introduction}
\label{introduction}
Autism Spectrum Disorder (ASD) is a complex neurodevelopmental disability with increasing prevalence.  In 2020, 1 in 36 children at the age of 8 was estimated to have ASD, higher than previous estimates during 2000-2018 \cite{maenner2023prevalence}. Patients exhibit symptoms such as social interaction impairments and communication deficits, causing challenges for them to adapt to society. This makes the early diagnosis of ASD very important. However, the symptoms of ASD are notably heterogeneous, including social, emotional, and cognitive impairments \cite{hernandez2015neural}. The lack of quantitative measures impedes the diagnosis process, calling for more advanced data analysis algorithms to uncover more accurate biomarkers and bolster timely detection and intervention.

% Recent research has increasingly focused on brain network analysis as a means of probing the underlying neurobiological mechanisms of ASD, with the goal of identifying network relationship patterns for brain activities. By analyzing brain signals extracted from functional Magnetic Resonance Imaging (fMRI) or electroencephalography (EEG) data, these studies aim to uncover the spatiotemporal dynamics of human brains and facilitate disease analysis. 

Brain network analysis has become a crucial approach for studying neurological disorders, leveraging data from functional Magnetic Resonance Imaging (fMRI) and other modalities to uncover the brain's spatiotemporal dynamics.
Specifically, temporal information has gained great emphasis. Compared with stationary data, dynamic signals can provide richer details of the brain \cite{campbell2023dyndepnet,lin2022deep,hutchison2013dynamic}. 
% Many disease patterns and natural phenomena can be uncovered by examining the time-varying brain signals. 
For example, Damaraju et al. \cite{damaraju2014dynamic} found that schizophrenia patients, on average, are less involved in large-scale brain activities, while they display more abnormal activities.

Despite these advances, one of the major challenges of studying brain activities is the immense amount of data involved, which awaits powerful algorithms to extract meaningful patterns and reduce the computation burden \cite{chung2019brain}. It is thus crucial to discern the underlying topology and structure of the data. 

% Topology is the study of shapes and structures. 
Topological data analysis (TDA) aims to study the shape of data. The Mapper approach \cite{singh2007topological} in TDA captures the shape and topology of the high-dimensional data samples using an efficient graph representation, depicting the data distribution across subjects or time. Applying the Mapper algorithm, Saggar et al. \cite{saggar2018towards} studied brain activities at rest and under different tasks and found important connectivity differences among brain states. Later, they utilized node degrees of Mapper graphs to uncover important transition hub states in resting state fMRI \cite{saggar2022precision}. Despite the impactful discoveries, the analysis was limited to healthy subjects, and the exploration of graph metrics remained basic. It is thus interesting to adopt the Mapper algorithm in the context of brain diseases and explore different metrics on Mapper graphs to uncover biomarkers of brain alterations. 

Graph signal processing (GSP) in graph theory \cite{ortega2018graph} extends conventional digital signal processing concepts to irregular graph domains. GSP metrics are powerful tools for brain network analysis. Huang et al. \cite{huang2016graph} decomposed brain signals to different smoothness levels using graph spectral operations and discovered the correspondence between brain network activities and graph frequencies. Preti et al. \cite{preti2023graph} leveraged the structure-function coupling implication of graph frequency components, linking brain functional activities with neural architectures through node- and edge-level graph metrics. However, these methods construct graphs based on brain regions while the time information is collapsed, hindering detailed analysis of brain state distributions across time. Regarding this, the Mapper algorithm has the clear advantage of not collapsing the temporal resolution. It is then reasonable to combine GSP with the Mapper algorithm to uncover the intricate brain dynamics across time.

% This study introduces a novel framework that combines topological data analysis (TDA) with graph signal processing (GSP) to analyze brain dynamics in ASD. 
This paper studies the brain activity distributions across time for autism and normal people. By employing the Mapper algorithm, a graph representation of the brain state distribution landscape can be created for each subject. Then, by studying the GSP and TDA properties of these graphs, the differences between the normal and autism groups are discerned. 

The main contribution and novelty of this paper are as follows:

\begin{itemize}
    \item To the best of the authors' knowledge, it is the first work that applies the Mapper algorithm to study patients with neurological disorders. % shall we make this claim?
    \item This work inventively combines GSP with TDA, applying a variety of GSP metrics (including node-, graph-level, and frequency domain metrics) and TDA quantities to examine brain state distribution networks.
    \item The group differences identified in the graph metrics reflect meaningful biological traits of altered brain activities induced by ASD. These findings attest to essential symptoms of ASD. % and have potential usage in ASD diagnosis
\end{itemize}

\section{Methodology}
Given the neuroimaging data, from which time-varying vectors can be extracted, we apply graph theory and TDA techniques to convert it into graph data for brain network analysis. Then, the networks are examined in terms of their structure, connectivity, and topology using various graph metrics. By comparing the metric results of normal and autistic groups, we study the differences in dynamic functional connectivity induced by autism.

\subsection{Mapper algorithm}\label{TDA}

In this study, we employ the concept of ‘network topology’ to depict the organization of brain activation modes over time. Unlike traditional network analyses focused on spatial brain region co-functioning, our approach uses the Mapper algorithm to capture the temporal distribution landscape of brain activation modes. 
% Each node in the Mapper network represents a group of similar brain states, while edges indicate overlaps between groups, revealing the overall structure of dynamic brain states rather than their spatial correlations.
The nodes in our brain state network represent clusters of similar activation patterns, with edges between nodes indicating temporal transitions, or overlaps, between these states. This topology captures the flow and diversity of brain dynamics, revealing how the brain explores various states and transitions over time.

\begin{figure}[htb]
    \centering
    \includegraphics[width=\linewidth]{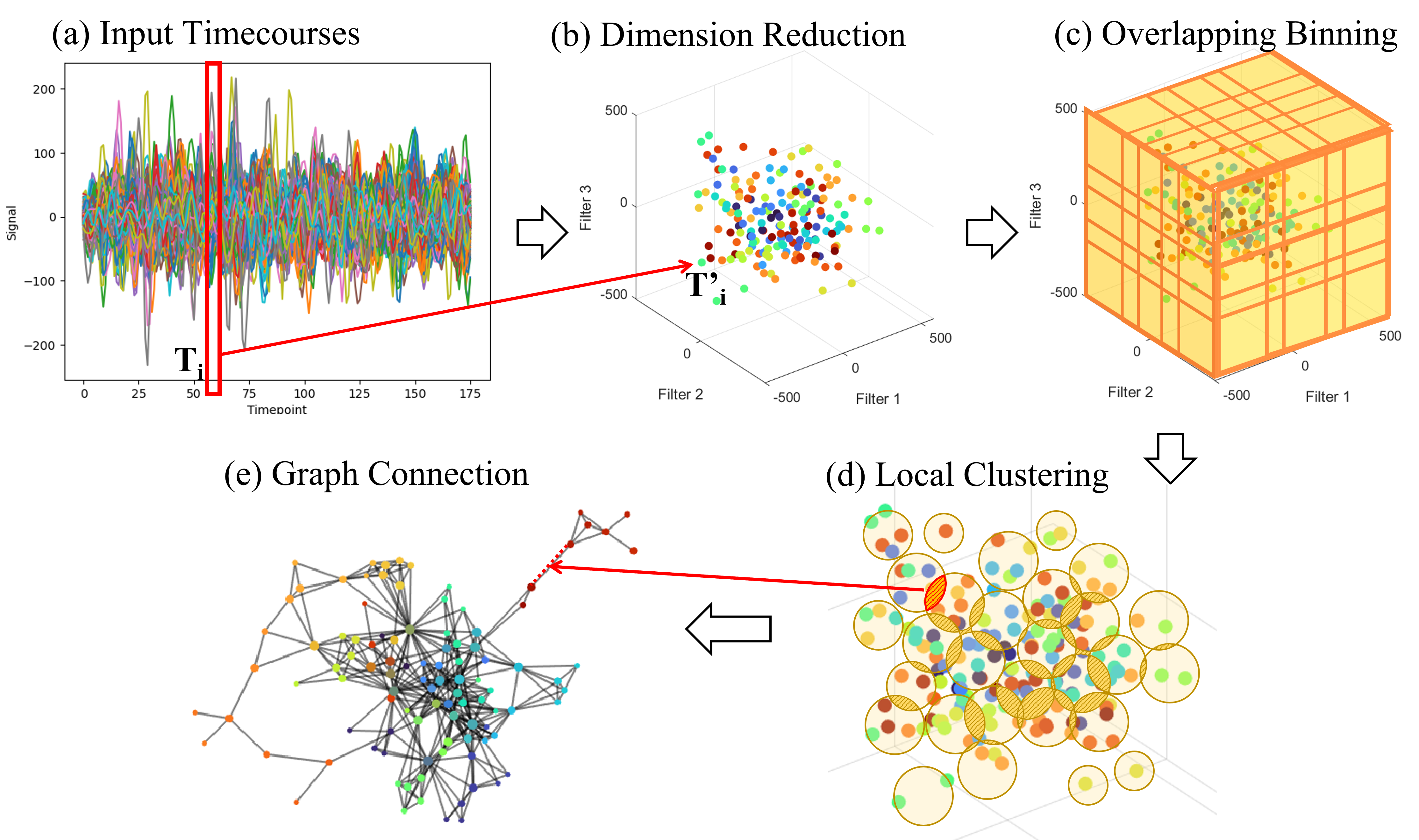}
    \caption{Diagram of the Mapper algorithm. (a) Extracted fMRI timecourses for all brain regions. (b) PCA projection of brain state instances. (c) Filter space divided into overlapping bins to enable clustering. (d) Clusters formed within each cube, represented as nodes based on similar brain states. (e) Nodes connected to form a graph based on shared data across overlapping regions. The final Mapper graph captures the topological relationships among dynamic brain states.}
    \label{mapper}
\end{figure}

% To reflect the brain state distribution at the individual level, we employ the Mapper algorithm from Topological Data Analysis (TDA). The Mapper algorithm encapsulates the topology of data samples as a graph structure, providing an essential depiction of the data distribution landscape. This approach makes few assumptions about the data and is robust to noise due to its use of the graph structure.

As illustrated in Fig.\ref{mapper}, the brain signal at each time point is viewed as a data vector in the data domain. The first step of the Mapper algorithm is to project data to a low-dimensional space where nearby points indicate similar brain activation patterns.  We use principal component analysis (PCA) to achieve this step. After that, the dimension-reduced data space is divided into overlapping bins, and partial clustering is conducted inside each bin to combine similar data samples (i.e., time frames). In the final step, a graph is constructed using the clusters as nodes and an edge is established between two nodes if the corresponding clusters have common data samples. 
% Through this process, we obtained a brain graph dataset, which enabled us to explore its topological and physical characteristics.
In the realization of Geniesse et al. \cite{geniesse2022neumapper}, which is adopted in our experiment, the graph is constructed as a reciprocal kNN graph, with an edge established between two nodes if they have common neighbors among the top-k nearest neighbors. This method stabilizes the noise level by reducing "false" connections.

% We firstly performed principal component analysis (PCA) for dimensionality reduction. Then, following the methodology proposed by Saggar et al. \cite{saggar2018towards}, we utilized binning operations to perform partial clustering and obtain the nodes of the graph. Subsequently, we established edges based on the existence of common data samples between clusters. Through this process, we obtained a brain graph dataset, which enabled us to explore its topological and physical characteristics.

Fig.~\ref{mapper} (e) shows a graph created by the Mapper algorithm. Through the binning process in the data sample space, the Mapper graph serves as a "histogram" of instantaneous brain states, depicting the distribution (probability) of a subject's brain activities across time. 

\begin{figure}
    \centering
    \includegraphics[width=0.8\linewidth]{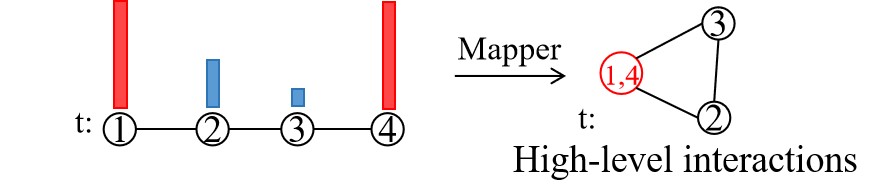}
    \caption{Formation of high-level temporal topology in the Mapper algorithm. Grouping similar signals in different time points enables the algorithm to model complex organizations in time.}
    \label{mapperstructure}
\end{figure}

The Mapper algorithm clusters similar time points into the same nodes, each representing a distinct brain activation mode. By connecting overlapping clusters, the resulting Mapper 
% graph captures transitions between different brain states, providing a topological map of brain state evolutions over time.
network encapsulates a complex topological map of brain state arrangements rather than a simple chain in chronological order, as illustrated in Fig. \ref{mapperstructure}. 
The topology of this network provides rich insights into the data distribution landscape. This approach makes few assumptions about the data and is robust to noise due to its use of the graph structure. 

% One important quality of the Mapper algorithm is that through the clustering process, similar time points can be merged as the same node, thus capturing more complex organizations rather than a plain chronical chain, as illustrated in Fig. \ref{mapperstructure}. 

\subsection{Distance correlation graphical model}

Distance correlation graphical models (DCGMs) are used in addition to the Mapper algorithm to model brain state networks. One of the most valued properties of DCGMs is that there is no PCA dimension reduction step, allowing the algorithm to treat every time point independently to keep the chronological information intact.
% and is vital to time-varying brain activity analysis.

\begin{figure}
    \centering
    \includegraphics[width=\linewidth]{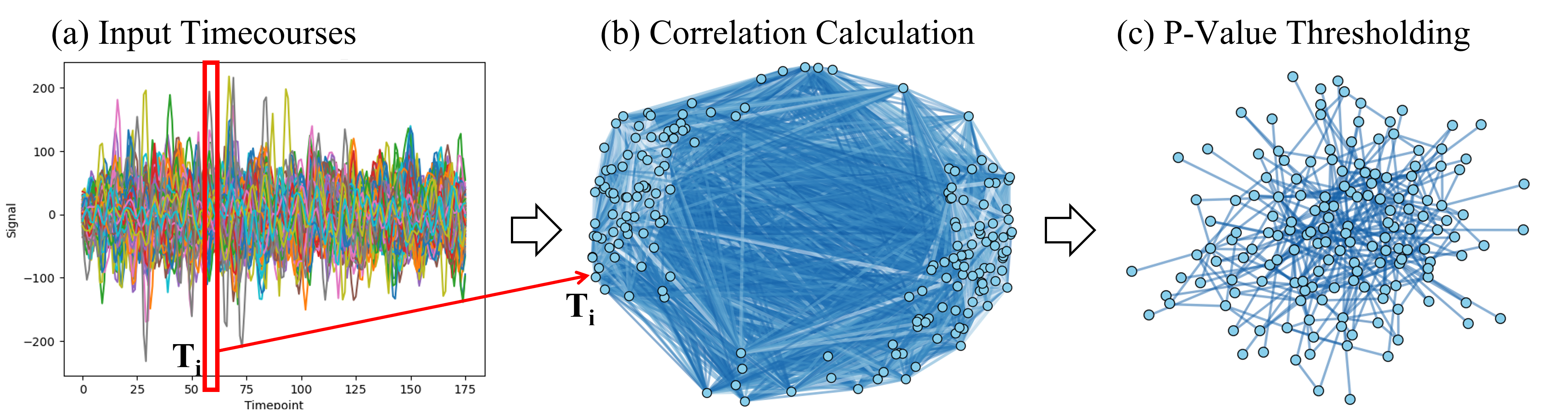}
    \caption{Diagram of the distance correlation graphical model.}
    \label{DCGM}
\end{figure}

As Fig. \ref{DCGM} shows, we construct the DCGM by calculating the distance correlation between each pair of time points. We then apply a p-value threshold to filter the correlation values, removing any edges with p-values greater than 0.05. This ensures that only statistically important connections are kept in the graph. In the resulting graph, each node represents a time point, and each edge indicates the correlation between the time points on its two ends.
% As shown in Fig. \ref{DCGM}, we construct the Distance Correlation Graph Model (DCGM) by calculating the distance correlation between each pair of time points. We then apply a p-value threshold to filter the correlation values, removing any edges with p-values greater than 0.05. This ensures that only statistically significant connections are retained in the graph. In the resulting graph, each node represents a time point, while each edge indicates the correlation between the two time points at its ends.

Distance correlation measures the association between random variables by comparing the pairwise distances in the joint distribution of variables to their marginal distributions. 
% Here's the mathematical formulation and some insights into DCGMs:
For two random variables $X$ and $Y$, the distance correlation between them is defined as follows:
\begin{equation}
    \text{dCor}(X, Y) = \sqrt{\frac{{\text{dCov}^2(X, Y)}}{{\text{dVar}(X)\text{dVar}(Y)}}}
\end{equation} 
where $dCov$ is the distance covariance, and $dVar$ is the square root of the product of distance variances.
% The distance covariance between two random vectors $X$ and $Y$ with dimensions \( n \) and \( m \) respectively is calculated as:

% \begin{equation}
%      \text{dCov}(X,Y) = \sqrt{\frac{1}{n^2} \sum_{i=1}^{n} \sum_{j=1}^{n} \delta_{ij} \delta_{ij}'}
% \end{equation}
% where \(\delta_{ij} = (X_i - X_j)(Y_i - Y_j)\), and \(n\) is the number of observations.
% The distance variance of a random vector X is defined as:
% \begin{equation}
%      \text{dVar}(X) = \sqrt{\frac{1}{n^2} \sum_{i=1}^{n} \sum_{j=1}^{n} \delta_{ij} \delta_{ij}'} 
% \end{equation}
% where \(\delta_{ij} = (X_i - \bar{X})(X_i - \bar{X})'\), and \(\bar{X}\) is the mean of the observations.

DCGMs capture nonlinear relationships between variables, which is a limitation of traditional linear methods like covariance or partial correlation. This allows DCGMs to model complex dependencies more accurately. Distance correlation is less sensitive to outliers compared to Pearson correlation, making DCGMs more robust in the presence of noisy data.

% Overall, Distance Correlation Graphical Models offer a flexible and powerful framework for modeling multivariate data, especially when the relationships between variables are nonlinear or when traditional linear methods are inadequate.

\subsection{Graph measures}
We introduce the graph modularity metric and frequency domain analysis of graph eigenvalue distributions to examine the Mapper and DCGM networks. These metrics are introduced below:

\textbf{Modularity}: Modularity assesses the presence of community structures within the network, identifying groups of nodes that exhibit higher connectivity with each other than with the rest of the network. This metric measures the degree to which a network can be divided into modules, which unveils the modular organization of the brain's temporal state arrangement. Its mathematical expression is:
   \begin{equation}
        Q = \frac{1}{2m} \sum_{ij} \left( A_{ij} - \frac{k_i k_j}{2m} \right) \delta(c_i, c_j)
   \end{equation}
where \( A_{ij} \) are elements of the adjacency matrix, \( k_i \) is the Degree of node \( i \), \( m \) is the total number of edges in the graph, \( c_i \) and \( c_j \) are Community assignments of nodes \( i \) and \( j \), and \( \delta(c_i, c_j) \) is the Kronecker delta, which is equal to 1 if \( c_i = c_j \) and 0 otherwise.

\textbf{Laplacian Eigenvalues}: Eigenvalue distributions of the graph's Laplacian matrix reflect the overall network dynamics. The mean or variance of Laplacian eigenvalues reflects stability and the presence of distinct functional modules.
We can obtain a Laplacian matrix $L$ by $L = D - W$, and the degree matrix $D \in \mathbb{R}^{n\times n}$ is a diagonal matrix with diagonal elements where $D[i, i] = \sum\limits_{j \neq i}W[i,j]$ represents the weighted degree of the nodes. It can be proved that $L$ is real, symmetric, and positive semi-definite. 
Hence, we can decompose $L$ using eigendecomposition $L = U\Lambda U^H$ such that the diagonal matrix $\Lambda$ has entries: 

\begin{equation}
\lambda_i=\{0=\lambda_0 \leqslant \lambda_1 \leqslant \ldots \leqslant \lambda_{n-1}\}
\end{equation}
which will be the Laplacian eigenvalues of the network. Generally, lower eigenvalues correspond to smoother signals across the graph, where there is less variation among node values, while higher eigenvalues indicate the signal components that change abruptly among nodes \cite{ortega2018graph}.

The average eigenvalue of the graph’s Laplacian matrix offers insight into the smoothness of brain state transitions. Lower values indicate smoother transitions, whereas higher values suggest abrupt changes. By examining eigenvalues, we gain a frequency-based view of the brain’s dynamics, helping to differentiate the more gradual shifts in control groups from potentially more abrupt changes in ASD.
% entropy

\textbf{Von Neumann entropy:} 
Von Neumann entropy is employed here to characterize the variability or "mixedness" of brain activation modes. 
% It reflects whether the brain's state experiences distinct modes or consistently in a specific mode (a 'pure' state, entropy = 0). 
Higher entropy values suggest more diverse brain activity configurations, while lower values imply stable or focused activation modes. We can characterize the degree of randomness of a graph structure as the von Neumann entropy of its Laplacian eigenvalues. The mathematical formulation is:
\begin{equation}
    S = - \sum_{\lambda_i \neq 0}{\frac{\lambda_i}{len(\lambda)} \times \log{\frac{\lambda_i}{len(\lambda)}}},
\end{equation}
    where $\lambda$ is the set of all eigenvalues, $\lambda_i$ is the $i^{th}$ eigenvalue, and $len(\lambda)$ represents the total number of Laplacian eigenvalues.

% Von Neumann entropy quantifies the diversity of brain states. Higher entropy values suggest a more mixed or variable state structure, while lower values imply more stable, recurring patterns. In ASD, higher entropy may indicate less organized or more scattered brain states, aligning with restricted and repetitive behaviors in ASD symptoms.

\subsection{Topological analysis}
To examine the brain state distributions of the subjects, we explore the structural aspect of the simplicial complex composed of instantaneous brain state data points. 

% \begin{itemize}
    \textbf{Betti number:}
    In topological analysis, a Vietoris-Rips complex can be constructed by selecting a distance threshold, building edges between data points that are less distant than the threshold, and creating higher-level topological structures wherever possible. Then, the numbers of high-dimensional shapes are counted as the characteristics of the simplicial complex, which is annotated as the Betti numbers.

\textbf{Ollivier-Ricci curvature:} Olliview-Ricci curvature \cite{ollivier2009ricci} describes the connectivity and topology of each node in the graph. For node $x$, $y$ $\in V$ in $G(V,E)$, their Ollivier-Ricci curvature can be defined as following:
\begin{equation}
    ORC(x, y)=1-\frac{W\left(m_x, m_y\right)}{d(x, y)}
\end{equation}
where $W\left(m_x, m_y\right)$ is the $L^1$ Wasserstein distance \cite{ni2015ricci} and $d(x, y)$ is the shortest path distance between $x$ and $y$. 

\section{Experiment and result}

\subsection{Data preparation}

% The experiment data includes the study of brain state distribution differences between autistic and normal people. By using the Nilearn package \cite{abraham2014machine} in Python, the pre-processed data in the ABIDE dataset \cite{craddock2013neuro} can be obtained. The pre-processed steps include bandpass filtering, global signal regression, and quality check, ensuring a higher data quality. The fMRI data is then registered and masked using the icbm152 2009 \cite{fonov2009unbiased} to extract time series. These pre-processing steps are crucial in brain data analysis, as factors like individual differences and motion noise can lead to data ambiguity. As is shown in Fig. \ref{comparison}, the pre-processing and quality of data can make a profound impact on the analysis in the later stage.

The neuroimaging data used in this study is from the New York University (NYU) subset of the preprocessed Autism Brain Imaging Data Exchange I (ABIDE I) dataset \cite{craddock2013neuro}. 
% The ABIDE I initiative provides open-access neuroimaging data to support the study of autism spectrum disorder (ASD). 
The data includes resting-state functional MRI (rs-fMRI) scans for individuals diagnosed with ASD and typically controls (TCs). Each participant’s data includes rs-fMRI sampled over approximately 200 time points with a total duration of six minutes.
The NYU subset includes 184 subjects, among which we use the data of 74 ASD and 98 TC subjects available in the Nilearn package \cite{abraham2014machine} in Python. These subjects aged from 6.5 to 39.1 years, among which approximately 80\% are male. 

\begin{figure}[h!]
    \centering
    \includegraphics[width=1\linewidth]{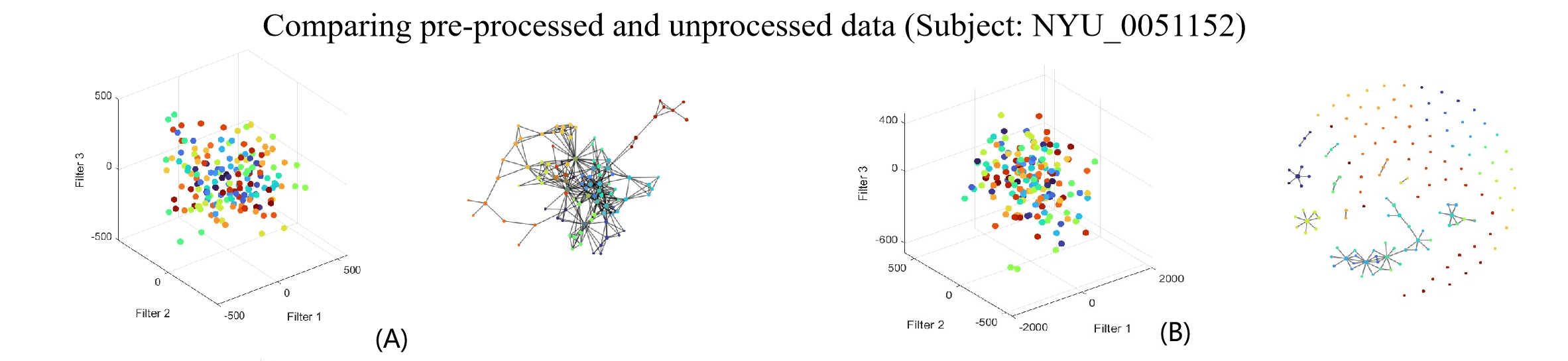}
    \caption{A comparison of graphs created with pre-processed and unprocessed data. (A). Graph created using data with bandpass filtering or global signal regression. (B). Without bandpass filtering or global signal regression. The parameters for graph creation in both subfigures are the same, as mentioned in Section~\ref{netcons}.}
    \label{comparison}
\end{figure}

The downloaded scans have been preprocessed with bandpass filtering, motion correction, spatial normalization, global signal regression, and quality check, ensuring the data quality. The fMRI files are then registered and masked using the icbm152 2009 template \cite{fonov2009unbiased} to extract timecourses from 360 regions of interest (ROIs) in floating point array format. These pre-processing steps are crucial in brain data analysis, as factors like individual differences and motion noise can lead to data ambiguity. As is shown in Fig. \ref{comparison}, the pre-processing and quality of data can make a profound impact on the analysis in the later stage.

\subsection{Network construction}
\label{netcons}% parameter selection, network illustration, more about the neighbors

To create the networks that reflect the brain state distribution at the individual level, we examine both the Mapper algorithm and the DCGM.

Fig. \ref{comparison} (A) shows an example of the graph created by the Mapper algorithm. Some parameters in the Mapper algorithm can affect the resulting graph
% , as introduced in \ref{themapperalgorithm}
.
After taking note of the parameter searching process of Saggar et al. \cite{saggar2018towards} and testing out different choices, we set the parameters as below in this experiment:

% (see Fig. \ref{param})
% The parameter searching is also guided by a few graph metrics.
% \begin{figure}
%     \centering
%     \includegraphics[width=\linewidth]{figures/mapper.png}
%     \caption{The parameter selection process.}
%     \label{param}
% \end{figure}

% list of parameter choices
\begin{itemize}
\itemsep0em
\topsep0pt
    \item Resolution ($R$): 50
    \item Gain ($G$): 30\%
    \item Number of neighbors ($k$): 6
    \item Dimension reduction method: PCA with 50 PCs
\end{itemize}

% in Fig. \ref{res}.
% \begin{figure}
%     \centering
%     \includegraphics[width=\linewidth]{figures/res_sub308.png}
%     \caption{accuracy with different numbers of principal components.}
%     \label{res}
% \end{figure}

For the construction of the DCGM, we first calculate the distance correlation between brain signal vectors at different time points. Then, we screen out the statistically insignificant ones using a p-value threshold of 0.05, thus obtaining the weighted adjacency matrix, as shown in Fig.~\ref{DCGM}.

% We plot the resulting DCGM matrices of some subjects in Fig. \ref{metastates}, where we can observe the metastate structures and brain state organizations in both autism patients and normal controls, in similar fashions as discovered in the previously mentioned paper\cite{vidaurre2017brain}. 

% \begin{figure}
%     \centering
%     \includegraphics[width=\linewidth]{figures/res_sub308.png}
%     \caption{accuracy with different numbers of principal components.}
%     \label{metastates}
% \end{figure}
% \cite{vidaurre2017brain}
\subsection{Graphical and topological metrics plots}

The metric findings and comparisons of the brain-state graphs generated by the Mapper algorithm and the DCGM are plotted in Fig.\ref{metric_results}. The violin plots illustrate the distributions of metric values in autistic and control groups, with box plots inside indicating the median and quartiles of the distributions. The histograms show the detailed distributional features of the two groups' metric values and their comparison. By observing these plots, including the peaks, tails, quartiles, and shapes of these distributions, we can make the following analysis of the data:

\begin{figure}[t!]
    \centering
    \includegraphics[width=\linewidth]{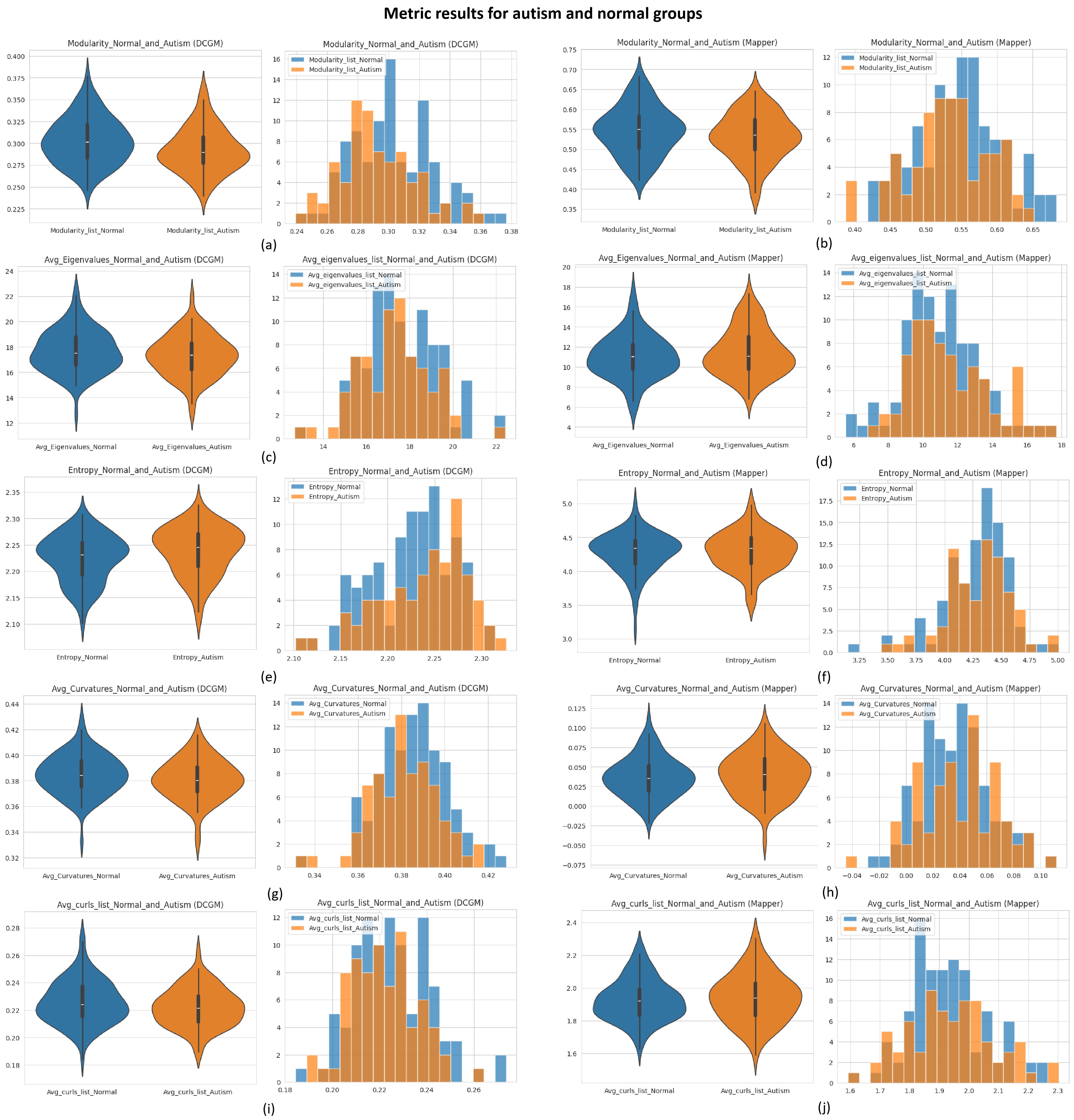}
    \caption{Results of graph and topological metrics for autism and normal groups. The left column (a, c, e, g, i) shows the violin plots and histograms of modularity, average eigenvalue, entropy, average curvatures, and average curls of DCGM graphs, while the right column (b, d, f, h, j) shows these metric results of the Mapper graphs. }
    \label{metric_results}
\end{figure}

\textbf{Modularity:}
    As shown in Fig. \ref{metric_results}(a-b), both graphs witness a tendency for decreased modularity in the autism group. These plots show that, compared with normal controls, autism patients tend to have lower network modularities on average, indicating less clear-cut community structures in the network. This contrast implies that ASD patients display less distinct types of brain activities or fewer transitions in brain states at rest. This result reflects that the autism subjects have less distinct "community" or "clusters" of brain states, which can indicate the overly smooth brain activity discovered in \cite{watanabe2017brain}.

    \textbf{Average eigenvalue:}
    The frequency domain analysis is conducted by studying the eigenvalue distribution, as indicated in Fig. \ref{metric_results}(c-d). Interestingly, the average eigenvalues of DCGM and Mapper graphs lead to contradicting results. But overall, there are no significant differences between the average eigenvalues of normal and autistic groups.
    
    % distinguish autism subjects clearly from normal controls, implying an underlying alteration of brain state distribution mechanisms of autism subjects. In general, autism subjects are likely to preserve lower average eigenvalues, and the average eigenvalues are less variant among the group of autism subjects compared with normal controls.
    
    According to the theory of GSP \cite{ortega2018graph}, eigenvectors of smaller eigenvalues correspond to smoother graph signals across nodes. However, comparing eigenvalues themselves may not have direct implications. Considering the temporal graph's ability to characterize brain state distributions across time, it can be valuable to examine the frequency components of temporal graph signals, as it can reveal brain dynamics over time. 
    % This corroborates the biological findings on autism that the patients have overly stable neural activities and lower speeds transitioning between different brain states.\cite{watanabe2017brain}

    \textbf{Von Neumann entropy:}
    As shown in Fig.~\ref{metric_results}(e), the ASD group exhibits higher Von Neumann entropy values, suggesting mixed brain activation states. This may correspond to the variable cognitive processes, restricted interests, and repetitive behaviors in ASD introduced in \cite{hernandez2015neural}. Lower entropy values in control subjects suggest a more stable or focused distribution of activation modes, indicating more consistent brain state patterns.
    
    % The networks of autism subjects are likely to have higher von Neumann entropy. The DCGM captures this phenomenon more clearly. The increased entropy values imply more random states of mind in autistic subjects at rest. Together with the comparison of network modularity, this shows that the autistic subjects have less organized, distinct brain activities at rest, which could be an indicator of restricted interests and repetitive behaviors as introduced in \cite{hernandez2015neural}. 

    \textbf{Node curvatures:}
    Fig. \ref{metric_results}(g-h) indicates the distributions of node curvature values in normal and autistic groups. The two graph creation methods produce very distinct curvature score distributions. Compared to DCGM graphs, Mapper graphs have much more variant node curvature values. This implies that the Mapper algorithm can create relatively more diverse graph structures.

    \textbf{Simplicial weights curl:}
    Simplicial weights curl depicts the flow of weights of the simplicial complex constructed from the brain network. According to Fig. \ref{metric_results}(i-j), the Mapper graph and the DCGM yield different curl comparisons for normal and autism groups. One possible explanation is that the simplicial weights curl is related to the methods of network creation. Hence, future work will consider simplicial weights curl as structural criteria to guide the network generation process and select the optimal network generation method that best reflects subjects' time-varying brain activity.

\begin{figure}[htbp]
    \centering
    \includegraphics[width=\linewidth]{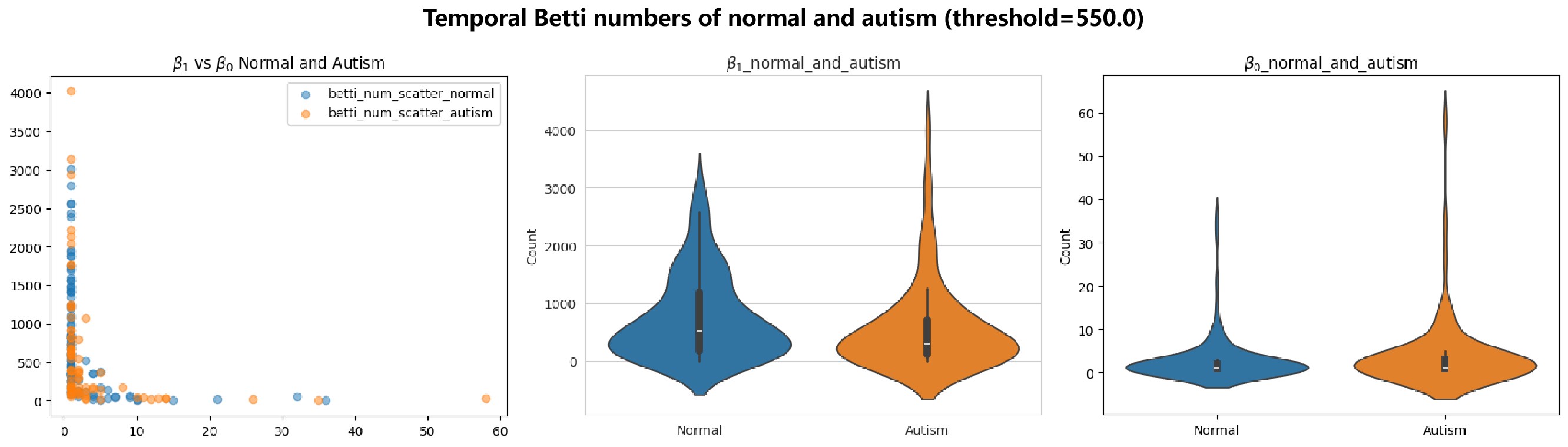}
    \caption{Betti numbers $\beta_0$ and $\beta_1$ for brain states of autism and normal subjects under the same threshold (distance $\leq$ 550). }
    \label{betti}
\end{figure}

\textbf{Betti numbers:}
In our experiment, the topological analysis is realized by the GUDHI package \cite{maria2014gudhi} in Python. The simplicial complex is created using a maximum edge length of 500. We collect the zeroth- and first-order Betti numbers of these patients, which are shown in Fig. \ref{betti}. As can be observed in the plots, compared to normal people, autistic subjects tend to have higher counts of $\beta_0$ numbers and decreased counts of $\beta_1$ numbers, indicating a larger number of scattered components and a lower amount of complex interaction structures. 

% individual experience
\textbf{Individual interpretations} The heterogeneity observed in the ASD group, as seen in metrics such as modularity and clustering, suggests that TDA-based methods are well-suited for capturing individual differences in brain networks. This variability aligns with the broad spectrum of ASD behavioral profiles and indicates potential for personalized insights into ASD-related brain function. 

\section{Conclusion}

This study used TDA and GSP metrics to investigate the difference in brain activities between ASD subjects and control individuals. By applying the Mapper algorithm and the DCGM, we constructed brain state distribution networks for each subject, enabling a deeper understanding of the underlying patterns in brain networks.

Both algorithms are accompanied by graph theory and topological data analysis to investigate the temporal pattern of brain connectivity. The application of temporal graphs provides a deeper understanding of the human brain's temporal transitions under various conditions. The results of graphical and topological metrics implicated less modularized and more variable brain states for autistic subjects, corroborating previous findings in neuroscience studies and proving the ability of graph and topological analysis to capture the alteration in brain activity of neurological disorders.

\bibliographystyle{splncs04}
\bibliography{mybibliography}





% LNCS Homepage, \url{http://www.springer.com/lncs}, last accessed 2023/10/25
% \end{thebibliography}
\end{document}